\documentclass{article}

\usepackage{arxiv}

\usepackage[utf8]{inputenc} 
\usepackage[T1]{fontenc}    
\usepackage[hidelinks]{hyperref}       
\usepackage{url}            
\usepackage{booktabs}       
\usepackage{amsfonts,amsmath}       
\usepackage{nicefrac}       
\usepackage{microtype}      
\usepackage{lipsum}
\usepackage{graphicx}
\usepackage{bm}
\graphicspath{ {./images/} }

\usepackage{subfigure}
\usepackage{caption}
\newcommand\blfootnote[1]{%
  \begingroup
  \renewcommand\thefootnote{}\footnote{#1}%
  \addtocounter{footnote}{-1}%
  \endgroup
}

\title{Data-Driven Calibration of Analytical Concrete Creep Models Considering Preloading Effects Using Gaussian Processes}

\author{
 Leonie Heller$^{\star}$ \\
  Bauhaus-Universität Weimar\\
  Weimar, Germany \\
  \And
   Christopher Taube\\
  Bauhaus-Universität Weimar\\
  Weimar, Germany \\
  \And
   Gledson Rodrigo Tondo\\
  Bauhaus-Universität Weimar\\
  Weimar, Germany \\
  \And
 Guido Morgenthal \\
  Bauhaus-Universität Weimar\\
  Weimar, Germany \\
}

\begin{document}
\maketitle
\begin{abstract}
The time-dependent deformation of concrete, particularly creep, remains a key challenge for reliable and material-efficient design. Experimental results show that tailored preloading – short-term loads exceeding the subsequent sustained load – can reduce both the magnitude and variability of creep strains which may be associated with beneficial microstructural changes. 

Building on these insights, this article employs Gaussian Process Regression (GPR) to calibrate analytical creep models, incorporating the effects of preloading intensity, timing, and concrete age into conventional predictions. The study pursues three main objectives: (i) calibrating a creep model using GPR based on experimental data, (ii) evaluating the impact of training data selection and preparation, and (iii) analysing model performance depending on the available experimental duration. 

The results demonstrate that GPR can improve model accuracy, quantify uncertainties, and support optimal test planning, while also enhancing understanding of preloading effects and contributing to more reliable and sustainable concrete creep predictions.

\end{abstract}

\keywords{concrete creep \and preloading \and predictive models \and gaussian process regression (GPR) \and physics-informed gaussian processes (PIGP) \and model calibration}

\section{Introduction}\blfootnote{IABSE Symposium Copenhagen 2026 - Bridging Advanced Technologies - Structural Innovation\\ \hspace*{5mm} Copenhagen, DK, 2026}

Accurately predicting the creep behaviour of concrete poses a significant challenge for the safe and material-efficient design of structures, due to the large number of influencing factors involved. While overestimating creep can lead to uneconomical designs characterized by oversized cross sections or unnecessary superelevations, underestimations may result in unforeseen large deformations and uncontrolled cracking. These effects are particularly critical in prestressed structures, where time-dependent losses of prestressing force are strongly affected by creep. 

A series of experiments conducted at Bauhaus-Universität Weimar, described in detail in~\cite{Taube.2025,Taube.2025b}, demonstrates that the creep behaviour of concrete can be positively influenced by tailored preloading applied at an early concrete age. Concrete specimens subjected to short-term loading prior to the long-term test exhibited both reduced creep strains and a smaller scatter in creep behaviour. These findings suggest that the initial short-term loading induces microstructural changes in the concrete, which in turn affect its long-term load-bearing behaviour.

Various analytical modelling approaches to describe the time-dependent development of creep over time have been proposed in literature and building codes, accounting for factors such as concrete composition, environmental influences and loading conditions through their model parameters. Previous investigations~\cite{Heller.2025} comparing analytical models of different mathematical forms have shown that the experimentally observed development of the creep coefficient can be well represented by the hyperbolic function specified in Eurocode 2~\cite{ec2}. 

As part of this study, a Gaussian Process (GP)-based approach is employed to determine the model parameters. GPs are considered a powerful tool for regression problems, as their Bayesian framework enables the prediction of both the expected outcome and the associated uncertainty. By combining GPs with physical models, the optimal values of the model parameters, their standard deviations, and correlations can be identified~\cite{Cross.2024}.

The aim of this article is to calibrate the EC 2 creep model using a physics-informed Gaussian process (PIGP) framework, with particular emphasis on the effects of tailored preloading, the selection and preparation of training data, and the duration of available test data. An initial sensitivity analysis is conducted to quantify the influence of individual parameters and to identify suitable model configurations, which are then calibrated using a PIGP approach and employed to analyse various influencing factors.

\section{Physics-informed Gaussian Process model} \label{sec:2}

\subsection{Model definition}

This work defines a machine learning model to probabilistically represent the creep development over time based on observed test data. For this purpose, the creep model is defined as Gaussian Process: 

\begin{equation}
    \bar{\varphi} \sim \mathcal{GP}(\mu(t), k(t,t')+\epsilon)
\end{equation}

with a time-dependent mean and covariance function. To account for the physical behaviour of creep in the optimisation, the mean function follows the creep formulation given in EC 2~\cite{ec2}:

\begin{equation}
    \mu(t)=\varphi(t),
\end{equation}

while the covariance is calculated as the sum of the physics-informed part in the form of a squared exponential kernel

\begin{equation}
    k(t,t')=\sigma_s^2 \cdot exp \left(-\frac{(t-t')^2}{2l^2}\right)
\end{equation}

and a noise part

\begin{equation}
    \epsilon = \sigma_n^2\delta_{tt'}.
\end{equation}

Here, $\sigma_s^2$ is the variance correlated with the experimental data, $l$ the length scale controlling the covariance smoothness, $\sigma_n^2$ the measurements noise variance, and $\delta_{tt'}$ the Kronecker delta function. Based on a given data set $\mathcal{D}=\{\mathbf{y,t}\}$ which contains the creep coefficient $\bm{y}$ measured at a defined point in time $\bm{t}$, the model is evaluated as

\begin{equation}
    \bar{\varphi} \sim \mathcal{GP}(\bm{\mu}, \bm{K} + \sigma_n^2\bm{I})
\end{equation}

where $\bm{\mu}=\varphi(t), \bm{K}=k(t,t')$ and $\bm{I}$ is the identity matrix. The evaluation of this model depends on both the free parameters of the physical creep model and on those of the covariance function, which are summarized in the parameter vector $\bm{\theta}$.

\subsection{Parameter identification}
Since the optimal values of the parameters are a priori unknown, prior models are defined by assuming realistic ranges for the creep parameters. For a given set of parameters the model likelihood can be calculated in closed form as

\begin{equation} \label{eq:likelihood}
    p(\bm{y}|\bm{t,\theta})= \frac{exp \left( -\frac{1}{2}(\bm{y}-\bm{\mu})^T(\bm{K}+\bm{I}\sigma_n^2)^{-1}(\bm{y}-\bm{{\mu}})\right)}{(2\pi)^{-N/2}\text{det}(\bm{K}+\bm{I}\sigma_n^2)^{-1/2}},
\end{equation}

where $N$ is the number of measurement points contained in $\mathcal{D}$. Applying Bayes’ theorem to the likelihood determined using Eq. \ref{eq:likelihood}, taking into account the model parameters, yields the posterior distribution 

\begin{equation}
    p(\bm{\theta}|\bm{y,t)}=\propto p(\bm{y}|\bm{t,\theta}) \prod_ip(\bm{\theta_i)}.
\end{equation}

This posterior distribution generally does not have a closed form and is therefore approximated using Markov chain Monte Carlo sampling with the Metropolis-Hastings algorithm.

\subsection{Prediction}
Once the model has been trained and the posterior distribution of the parameters determined, the PIGP model can be employed to predict creep coefficients at time points without measurement data. Based on the training examples and the estimated optimal parameters, the creep coefficient at a new time point is predicted as a normal distribution:

\begin{equation}
p(\varphi (t_{\star}) | t_{\star}, \bm{y}, \bm{t}, \bar{\bm{\theta}} ) = \mathcal{N} \left( \mu_{\star}, \sigma_{\star}^2 \right),
\end{equation}

where the analytical solutions for the mean $\mu_{\star}$ and the covariance $\sigma_{\star}^2$ at time $t_{\star}$ are given by

\begin{align}
\begin{split}
 \mu_{\star} & = k(t_{\star},\bm{t}) \left( \bm{K} + \sigma_n^2 \right)^{-1} \bm{y},\\  
 \sigma_{\star}^2 & = k(t_{\star}, t_{\star}) - k(t_{\star},\bm{t}) \left( \bm{K} + \sigma_n^2 \right)^{-1} k(\bm{t}, t_{\star}).
 \end{split}
\end{align}

The posterior model generated using the Markov chain method allows the probability distributions of the model parameters to be taken into account, which in turn influence the prediction uncertainty of the model. This can be included in the creep predictions by integrating the parameter models

\begin{equation}
p(\varphi(t_{\star}) | t_{\star}, \bm{y}, \bm{t} ) = \int p(\varphi(t_{\star}) | t_{\star}, \bm{y}, \bm{t}, \bm{\theta} ) p(\bm{\theta} | \bm{y}, \bm{t}) d \bm{\theta}.
\end{equation}

This integration cannot usually be solved analytically and is therefore approximated using the Monte Carlo method

\begin{equation}
p(\varphi(t_{\star}) | t_{\star}, \bm{y}, \bm{t} ) \approx \frac{1}{N} \sum_{n=1}^N p(\varphi(t_{\star}) | t_{\star}, \bm{y}, \bm{t}, \bm{\theta}_n ),
\end{equation}

where $\bm{\theta}_{n}$ denotes samples from the posterior model $p\left(\bm{\theta} | \bm{y},\bm{t}\right)$.

\section{Calibration of EC 2 creep model variants}
\subsection{Sensitivity analysis and model selection}
According to Eurocode 2~\cite{ec2}, the creep coefficient is obtained by multiplying a final value $\varphi_0$ with a term describing its time-dependent progression:

\begin{equation}
    \varphi(t,t_0)=\varphi_0 \cdot \left( \frac{t-t_0}{\beta_H+t-t_0} \right)^n.
\end{equation}

The notional creep coefficient $\varphi_0$ and the progression parameter $\beta_H$ are determined based on ambient relative humidity, compressive concrete strength, concrete composition and component geometry. The final value approached by the creep coefficient is given by

\begin{equation}
    \varphi_0=\varphi_{RH} \cdot \beta_{f_{cm}} \cdot \beta_{t_0}
\end{equation}

as the product of the parameters

\begin{equation}
    \varphi_{RH}= \left(1+ \left( \frac{1-RH/100}{0,1 \cdot h_0^{1/3}} \right) \cdot \alpha_1 \right) \cdot \alpha_2,
\end{equation}

\begin{equation}
    \beta_{f_{cm}}= \frac{16,8}{\sqrt{f_{cm}}},
\end{equation}

and

\begin{equation}
    \beta_{t_0}= \frac{1}{0,1+t_{0,eff}^{0,2}}.
\end{equation}

In addition to the exponent $n$ that is defined as a constant value of 0,3, the function's development is controlled by the parameter $\beta_H$:

\begin{equation}
    \beta_H = 1,5 \cdot (1 + (0,012 \cdot RH)^{18}) \cdot h_0 + 250 \cdot \alpha_3.
\end{equation}



\begin{figure}[b!]
    \centering

    \begin{minipage}[b]{0.40\textwidth}
        \vspace{0pt}
        \centering
        
        \renewcommand{\arraystretch}{1.3}
        \begin{tabular}{c c c c}
            \hline
             & $t_{0,eff}$ & $h_0$ & $n$ \\
            \hline
            $\mu$ & 32,5 & 50,0 & 0,30 \\
            \hline
            $\sigma_s$ & 10 \% & 10 \% & 3 \% \\
            \hline
        \end{tabular}
        \renewcommand{\arraystretch}{1}

        \vspace{2.8cm}

        \captionsetup{skip=8pt}
        \captionof{table}{Parameter means and standard deviations used for global sensitivity analysis.\\
        }
        \label{tab:params_sensitivity}
    \end{minipage}
    \hfill
    \begin{minipage}[b]{0.50\textwidth}
        \centering
        \vspace{0pt}

        \includegraphics[width=\linewidth]{images/sobol_ec2_paper.png}

        \caption{First-order and total-order Sobol indices of the EC 2 model parameters ($t_{0,eff}, h_0, n$) depending on the creep duration.}
        \label{fig:sensitivity}
    \end{minipage}

\end{figure}

The calibration of the EC 2 approach is based on the creep test data from experiments described in~\cite{Taube.2025}, which were conducted at Bauhaus-Universität Weimar. Due to the constant climatic conditions maintained throughout the entire testing period, relative humidity (RH) and temperature (T) can be assumed fixed and are therefore not considered in the model calibration. Compressive strength tests performed after the creep experiments show similar strengths across all specimens, indicating that the concrete strength is unaffected by the level of preloading. Consequently, the concrete compressive strength $f_{cm}$, used directly and indirectly (via the parameters $\alpha$) in the calculation of the creep coefficient, is taken as the mean of the measured values.

The conducted tests indicate a reduced creep tendency in preloaded specimens. This suggests that early-age preloading may cause a mechanically induced densification of the microstructure, involving dissolution and reorganization of solid-hydrate phases and resulting in an enhanced ageing effect~\cite{Taube.2025}. Within the EC 2 model framework, this effect can be considered by increasing the parameter $t_{0,eff}$, which defines an effective concrete age derived from the actual age at the start of the creep tests and the specific concrete composition. Furthermore, the resulting densification may slow internal transport processes, which can be captured in the model by adjusting the effective thickness $h_0$. In the standard Eurocode 2 formulation, the exponent $n$ is defined as a constant, independent of environmental and material parameters. Due to its strong influence on the initial development of the creep function, $n$ is subsequently treated as a free model parameter for calibration.

The influence of the parameters on the behaviour of the creep model is assessed using a global sensitivity analysis that evaluates the Sobol indices. In this approach, the variance of the model output is decomposed according to the influence of the individual parameters. The first-order index represents the proportion of the total variance that is exclusively caused by the main effect of the respective parameter. In contrast, the total index accounts for the variance contribution resulting from both the direct effect of the parameter and its interactions with other model parameters~\cite{Saltelli.2008}. Table \ref{tab:params_sensitivity} lists the input values and standard deviations, which were chosen based on EC 2 guidelines and the observed standard deviations during calibration. The corresponding results, depending on the load duration, are shown in Figure \ref{fig:sensitivity}. The agreement of the Sobol indices obtained from the sensitivity analysis indicates that the overall variance of the EC 2 model is affected by the variance of the individual model parameters, but not by their interactions. Furthermore, a similar level of influence can be attributed to the parameters $t_{0,eff}$ and $h_0$, whereas $n$ primarily affects the creep development during the early stages of the test period. 

Building upon these insights, versions of the EC 2-based creep model with varying complexity are calibrated using the described physics-informed Gaussian Process framework. The prior models of the individual parameters used in the following analysis are defined as follows: 
\begin{equation}
\label{eq:GPModelPrior}
p(\bm{\theta}) = \left\{
        \begin{array}{@{}ll@{}}
        \mathcal{U} (0, 100),    & \text{\hspace{1cm}for } t_{0,eff},\\
        \mathcal{U} (10, 500),    & \text{\hspace{1cm}for } h_0,\\
        \mathcal{U} (0,2, 0,5),         & \text{\hspace{1cm}for } n,\\
        \mathcal{U} (0, 10),     & \text{\hspace{1cm}for } \sigma_n,\\
        \mathcal{U} (0, 100),    & \text{\hspace{1cm}for } \sigma_s, \\
        \mathcal{U} (0, 1000),    & \text{\hspace{1cm}for } l.
        \end{array}
        \right.\
\end{equation}

The investigations include full three-parameter configurations as well as reduced two- and one-parameter versions, with any non-calibrated parameters fixed at their EC 2-prescribed values. 

\subsection{Three-parameter approach}
\begin{figure}[b!]
\centering
\begin{minipage}[t!]{0.49\linewidth}
    \includegraphics[width=\linewidth]{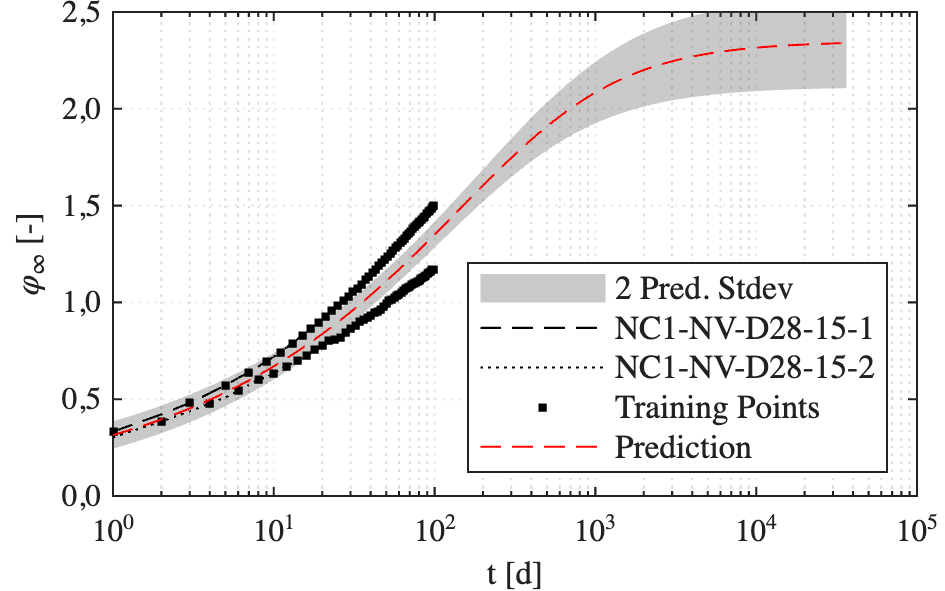}
\end{minipage}%
\begin{minipage}[t!]{0.49\linewidth}
    \includegraphics[width=\linewidth]{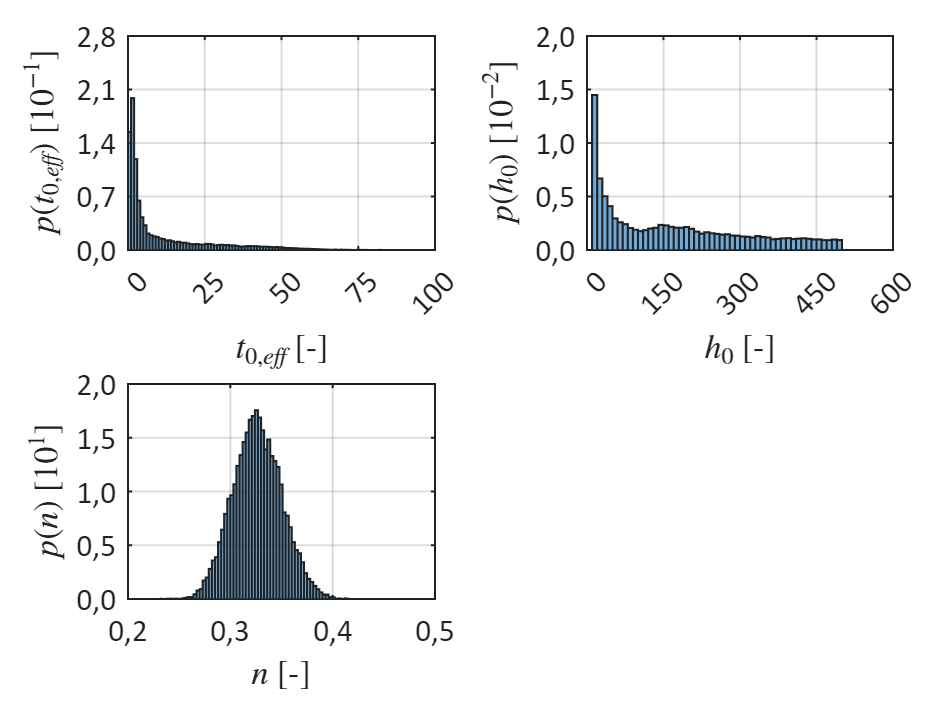}
\end{minipage}
\caption{Predicted creep coefficient (left) and distribution of optimised model parameters (right) based on the test data of the non-preloaded specimens using the three-parameter approach.}
\label{fig:3paramApproach}
\end{figure}

The optimisation results of the EC 2-based model, characterized by $\theta=\{t_{0,eff}, h_0, n, \sigma_n, \sigma_s, l \}$, are shown in Figure \ref{fig:3paramApproach} for the non-preloaded specimens. The predicted creep development (left) fits the experimental data well, accurately reproducing the observed creep progression. However, the distributions of the model parameters shown in Figure \ref{fig:3paramApproach} (right) reveal that both $t_{0,eff}$ and $h_0$ are calibrated towards their lower bound of the prior model, leading to optimised values that deviate significantly from those suggested in EC 2. This may be attributed to the model’s similar sensitivity to both parameters, which complicates their distinct identification.

\subsection{Two-parameter approach}
\begin{figure}[h]
\centering
\begin{minipage}[t!]{0.49\linewidth}
    \includegraphics[width=\linewidth]{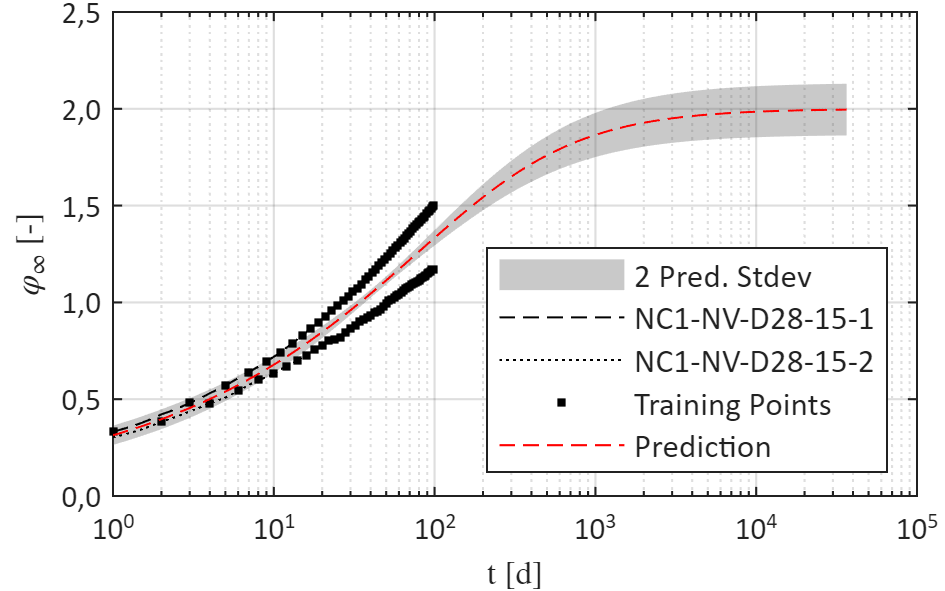}
\end{minipage}%
\begin{minipage}[t!]{0.49\linewidth}
    \includegraphics[width=\linewidth]{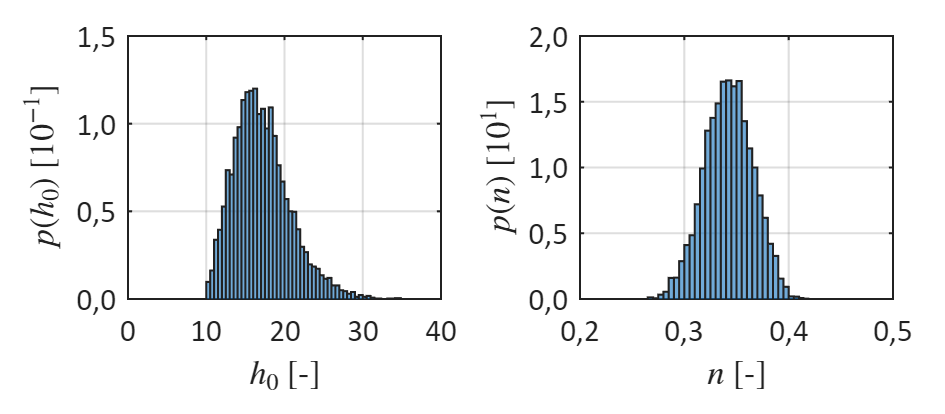}
\end{minipage}
\caption{Predicted creep coefficient (left) and distribution of optimised model parameters (right) based on the test data of the non-preloaded specimens using the two-parameter approach.}
\label{fig:2paramApproach}
\end{figure}

\begin{figure}[h]
\centering
\includegraphics[width=0.45\linewidth]{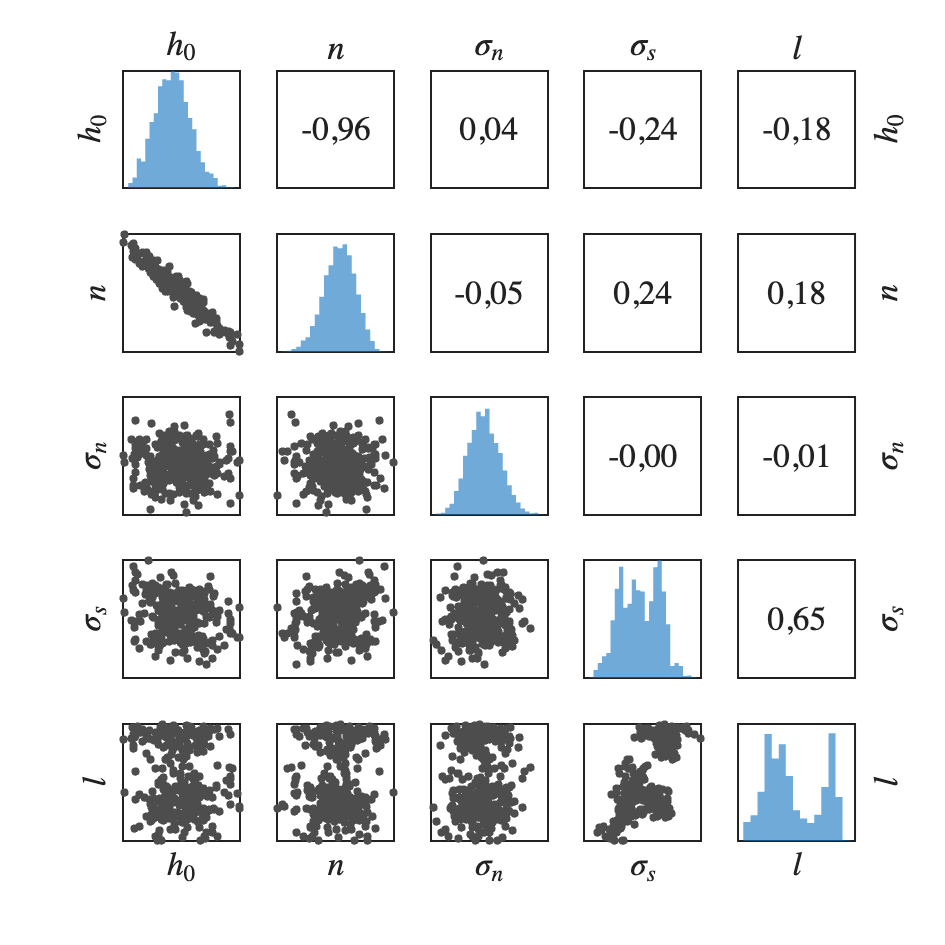}
\caption{Parameter correlations of the creep model and the hyperparameters of the covariance kernel.}
\label{fig:correlations}
\end{figure}

In order to achieve clear identification of model parameters and avoid the calibration issues observed, a two-parameter PIGP model defined by the parameter vector $\theta=\left\{h_0,n,\sigma_n,\sigma_s,l\right\}$ is employed, which describes the time-dependent creep behaviour as a function of $h_0$ and $n$. In contrast to $t_{0,eff}$, $h_0$ influences the notional creep coefficient $\varphi_0$ as well as the progression parameter $\beta_H$, since it is involved in the calculation of both. Figure \ref{fig:2paramApproach} (left) illustrates the predicted creep coefficient for the non-preloaded specimens, showing excellent agreement with the experimentally measured creep curves. Furthermore, the predicted creep coefficient and its associated standard deviation are lower than those obtained from the previous calibration. Regarding the optimised model parameters, depicted in Figure \ref{fig:2paramApproach} (right), stable distributions and values within physically reasonable ranges can be observed. The results shown in Figure \ref{fig:correlations}, however, reveal a strong correlation between $h_0$ and $n$, making it difficult to calibrate these two parameters independently.

\subsection{One-parameter approach}
\begin{figure}[t!]
\centering
\includegraphics[]{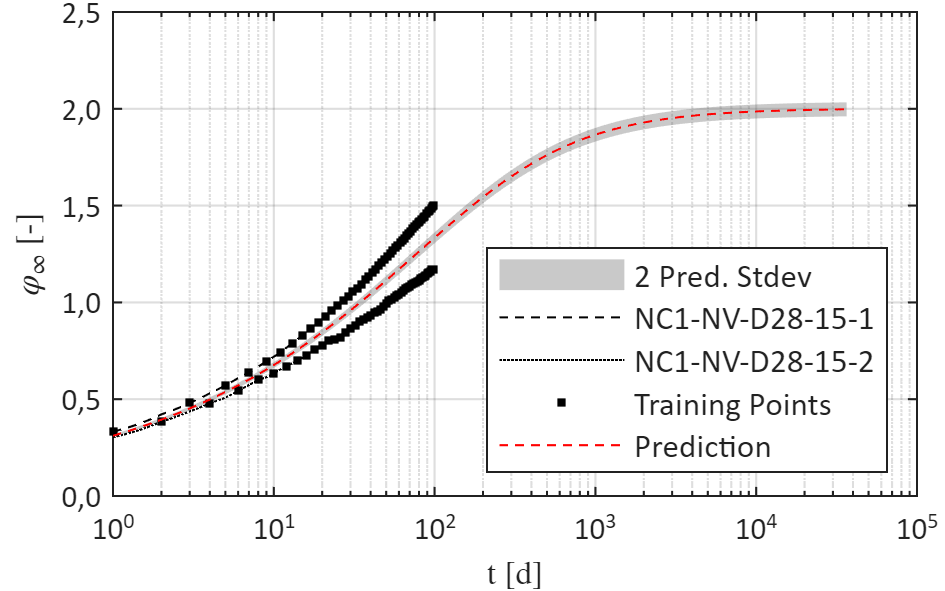}
\caption{Predicted creep coefficient with one parameter $h_0$ using the test data of the non-preloaded specimens and adjusted parameter $n$.}
\label{fig:1paramapproach}
\end{figure}

To avoid correlations between parameters during calibration, the PIGP model is reduced to a single parameter for calibration. The calibration of the three- and two-parameter approaches yields a similar value of n for all investigated models, which is adopted for the subsequent calibration of the single-parameter approach. Optimising the model characterized by $\theta=\left\{h_0,\sigma_n,\sigma_s,l\right\}$, using a value of 0,34 instead of the EC 2-prescribed value of 0,3, results in the creep function prediction shown in Figure \ref{fig:1paramapproach}. These results indicate that the single-parameter model provides prediction quality similar to that of the more complex models, while considerably reducing the associated standard deviation.

\begin{figure}[b!]
    \centering
    \includegraphics[width=\textwidth]{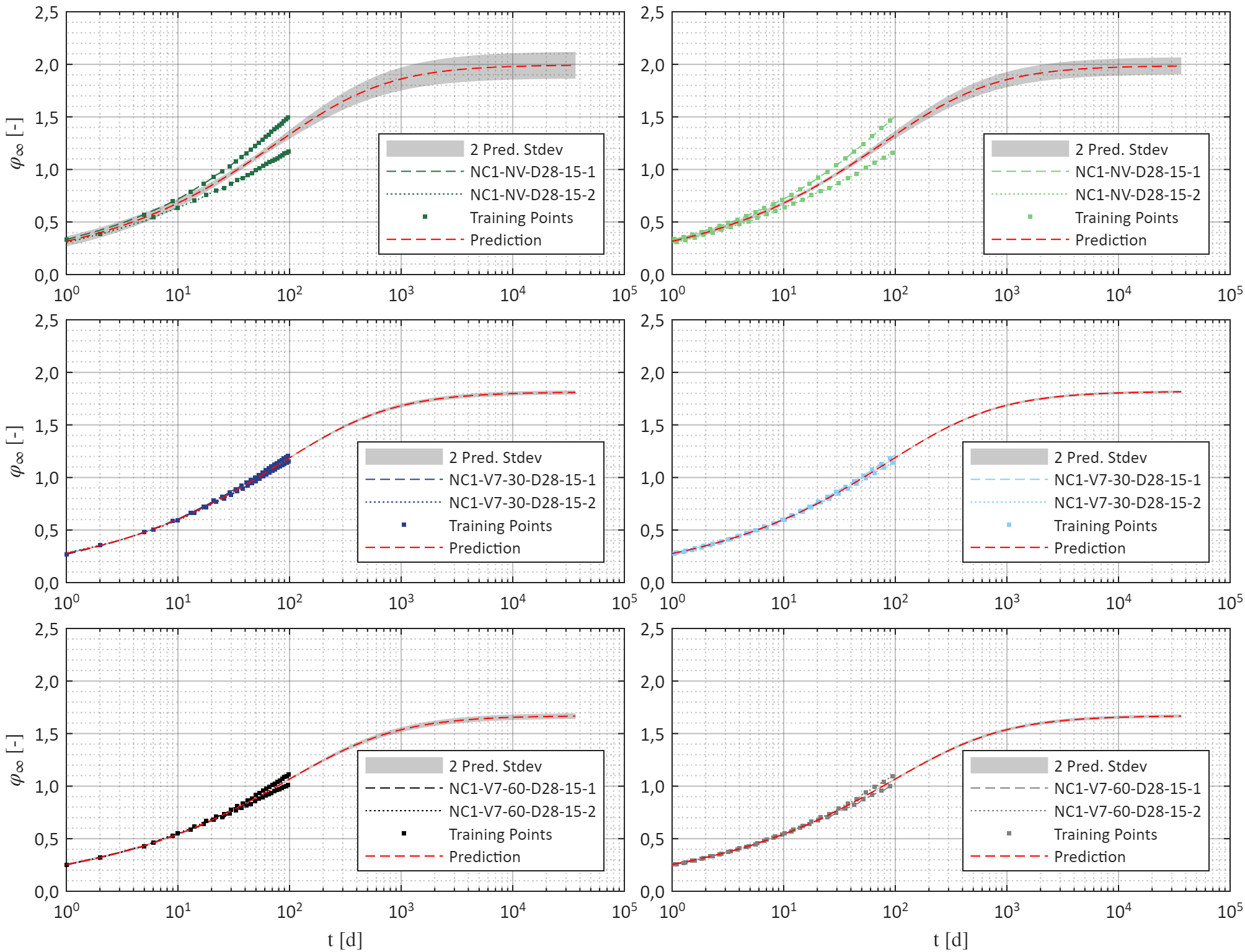}
    \caption{Predicted creep function for equidistant (left) and logarithmic (right) training data and 0 \% (green), 30 \% (blue) and 60 \% (black) preload intensity.}
    \label{fig:structuredata}
\end{figure}

\section{Investigation of influencing factors in the optimisation process}

Given the strong agreement between the predicted creep function and the experimental data, as well as the stable distributions of the calibrated parameters, the PIGP model based on $h_0$ and $n$ is employed for further investigations. The calibration is based on the creep tests from~\cite{Taube.2025}, including data from six specimens, two each preloaded to 60 \% and 30 \% at a young concrete age, as well as two non-preloaded specimens. After a stress-free hardening phase, creep loading was applied and sustained for a duration of 100 days. The model is optimised using datasets of 100 values each, distributed either equally or logarithmically over the test duration.

\subsection{Structure of training data}
Figure \ref{fig:structuredata} shows the predicted creep behaviour over a period of 100 years, obtained from model training with either equally spaced or logarithmically sampled datasets. The creep curves of the non-preloaded specimens (NC1-NV-D28-15) and those preloaded to 30 \% (NC1-V7-30-D28-15) and 60 \% (NC1-V7-60-D28-15) of their compressive strength are displayed as coloured training points, with individual models calibrated for each preload intensity. 
In contrast to the uniformly distributed data points across the 100-day test duration in the equally spaced dataset, logarithmic sampling provides a denser representation of the initial phase, thereby giving greater weight to the early creep development. Training with logarithmically sampled data results in a noticeable reduction in the standard deviation, while no significant differences are observed in the predicted creep coefficients.

\begin{figure}[b!]
\centering
\includegraphics[]{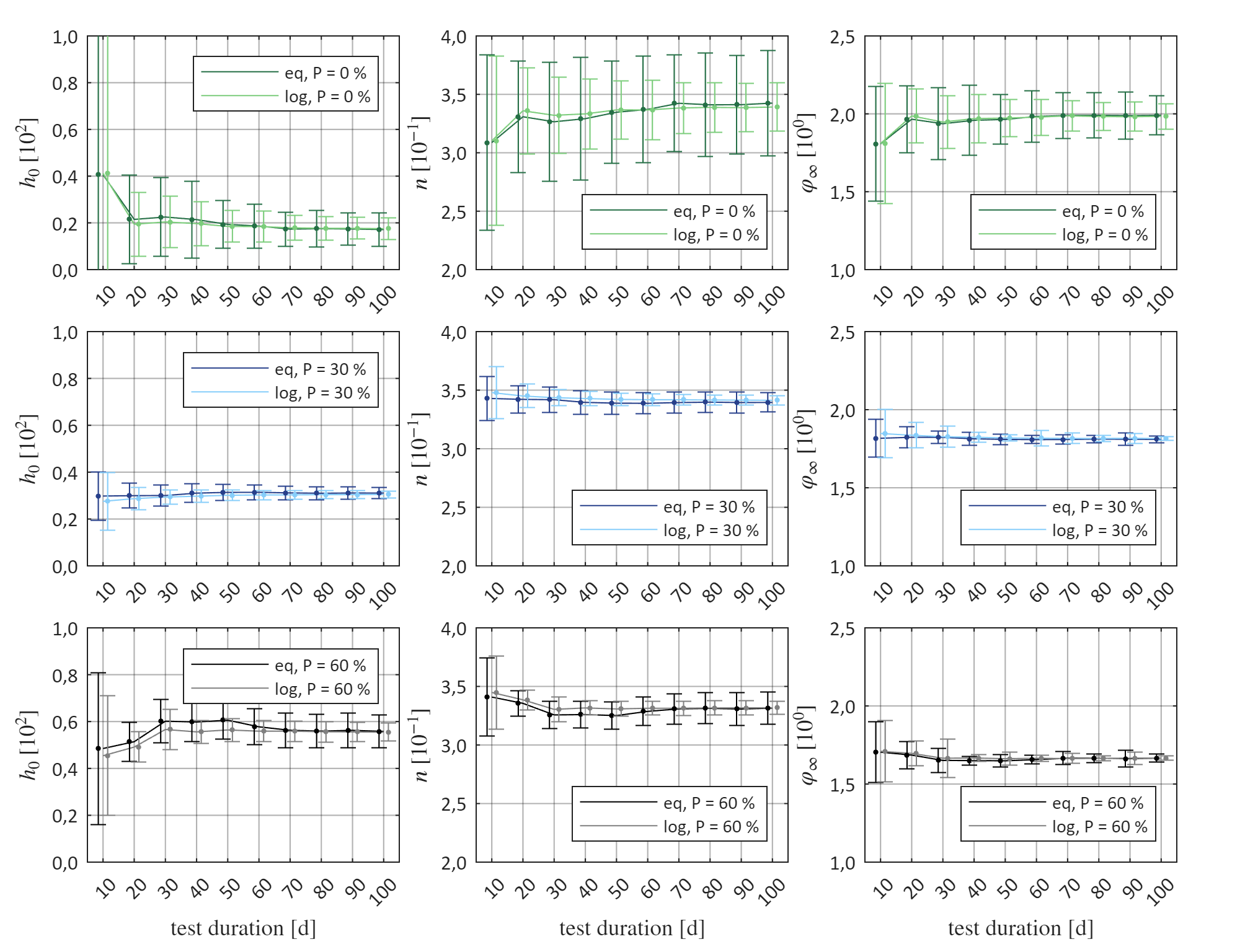}
\caption{Predicted creep function for equidistant (left) and logarithmic (right) training data and 0 \% (green), 30 \% (blue) and 60 \% (black) preload intensity.}
\label{fig:trainingduration}
\end{figure}

\subsection{Test duration}
Figure \ref{fig:trainingduration} illustrates the influence of test duration on parameter optimisation for specimens tested at different preloading levels, to determine the minimum of creep test duration required for robust model training. For parameters calibrated using logarithmically sampled data, lower standard deviations are obtained compared to those calibrated with equally spaced datasets. Regardless of the preloading level, an increase in test duration leads to a reduction in standard deviation, which remains nearly constant beyond approximately 60 days. Consistent with the scatter observed in the experimental data, the lowest standard deviations are found for specimens preloaded to 30 \% of the compressive strength, whereas higher deviations occur for tests with 60 \% preloading or without preloading. Similarly, the calibrated parameter values themselves become stable after a test duration of about 60 days.

\subsection{Preload intensity}
\begin{figure}[t!]
\centering
\begin{minipage}[t!]{0.49\linewidth}
    \includegraphics[width=\linewidth]{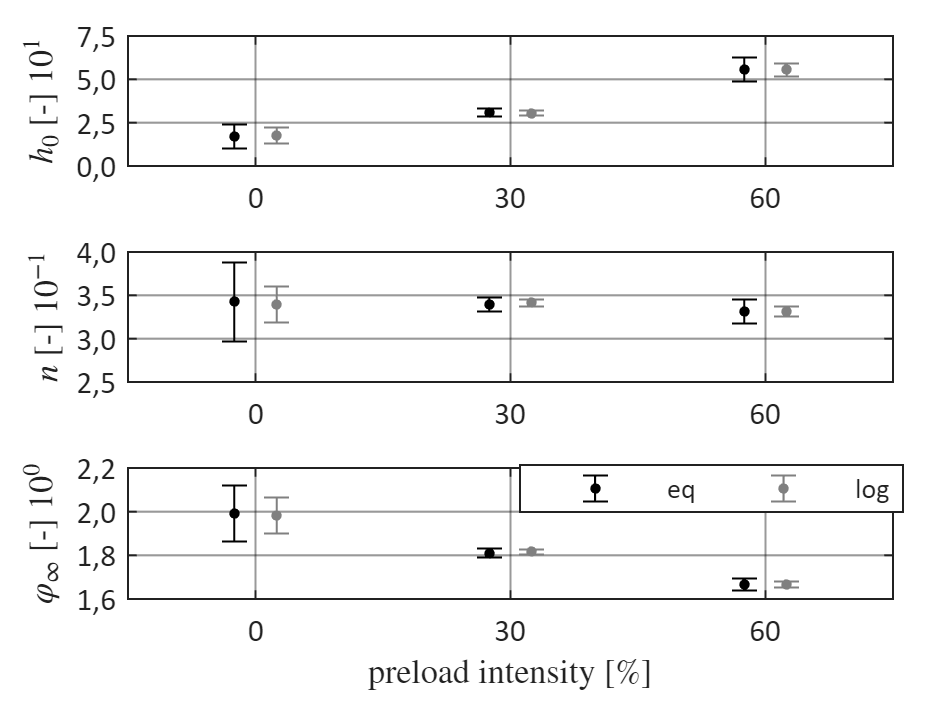}
\end{minipage}%
\begin{minipage}[t!]{0.49\linewidth}
    \includegraphics[width=\linewidth]{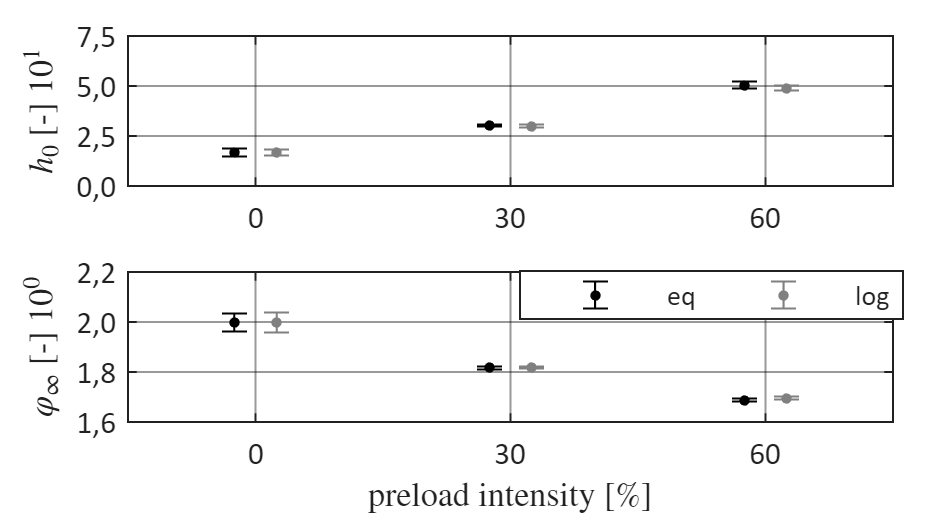}
\end{minipage}
\caption{Influence of preload intensity on calibrated model parameters and prediction of the final creep coefficient for two-parameter approach (left) and for one-parameter approach with adjusted $n$ (right).}
\label{fig:preloadintensity}
\end{figure}

The influence of the preload intensity on the calibrated parameters and the predicted final creep coefficient is shown in Figure \ref{fig:preloadintensity} (left). Higher preloading increases $h_0$ while simultaneously lowering the final creep coefficient $\varphi_\infty$, which supports the hypothesis that preloading induces microstructural densification and consequently slows internal transport processes. For a preloading level of 60 \% a decrease of approximately 16 \% can be observed. Across all preloading intensities, the exponent $n$ is calibrated to around 0,34, indicating that it is independent of preloading. Consequently, the approach can be simplified to a single-parameter model dependent only on $h_0$, as proposed in 3.4. The influence of the preload intensity on the model with $n$ fixed to 0,34, focusing on the calibrated parameter $h_0$ and the predicted creep coefficient $\varphi_\infty$ is shown in Figure \ref{fig:preloadintensity} (right). Compared to the two-parameter model shown in Figure \ref{fig:preloadintensity} (left), $h_0$ and $\varphi_\infty$ differ only slightly, while fixing $n$ leads to a significant reduction in the corresponding standard deviations.

\section{Conclusion}
The conducted investigations show that the creep coefficient trends observed in the experiments can be well captured using the selected optimisation parameters of the EC 2 model. From this, the following key findings can be derived:
\begin{itemize}
    \item Calibrating a model dependent on $h_0$ and n achieves the closest agreement with the creep test curves, but exhibits substantial parameter correlations.
    \item Logarithmically structured datasets yield lower standard deviations than those with equally spaced time steps.
    \item No significant changes in the optimised parameters occur beyond a test duration of approximately 60 days, indicating that future tests could be shortened without loss of model training accuracy.
    \item The independence of $n$ from the preload intensity implies that it is governed by a constant influencing factor, possibly related to the material composition.
    \item Fixing $n$ at the optimised value of 0,34 results in a correlation-free model that delivers comparable results to the $h_0$- and $n$-depending model while reducing prediction uncertainty.
\end{itemize}

The theoretical investigations presented in this article are based on a small experimental data base and should be validated through additional tests building upon the insights obtained. The analysis of different model variants indicates that the observed changes in creep behaviour due to preloading can be represented by adjusting the geometry parameter $h_0$, which characterises internal transport processes in concrete. This finding supports the hypothesis that preloading at an early concrete age may induce microstructural changes that slow down internal transport mechanisms, thereby possibly leading to a beneficial change in creep behaviour. 

The use of physics-informed Gaussian Processes has shown great potential in reproducing the time dependent creep behaviour and estimating the underlying model uncertainty. However, additional experimental data are required to enhance the robustness of the PIGP model, to deepen the understanding of concrete creep behaviour and enable realistic, predictive modelling.

\bibliographystyle{unsrt}  
\bibliography{references}

@misc{ec2,
title = {Eurocode 2: Design of concrete structures - {P}art 1-1: General rules and rules for buildings; {G}erman version {EN}1992-1-1:2004 + {AC}:2010},
 address = {Berlin},
 publisher = {{DIN Media GmbH}},
 doi = {10.31030/1723945}
}

@article{Cross.2024,
 author = {Cross, E. J. and Rogers, T. J. and Pitchforth, D. J. and Gibson, S. J. and Zhang, S. and Jones, M. R.},
 year = {2024},
 title = {A spectrum of physics-informed Gaussian processes for regression in engineering},
 volume = {5},
 journal = {Data-Centric Engineering},
 doi = {10.1017/dce.2024.2}
}

@inproceedings{Heller.2025,
 author = {Heller, L. and Taube, C. and Morgenthal, G.},
 title = {Analytical Modelling for the Prediction of the Improved Creep Behaviour of Tailored Preloaded Concrete},
 pages = {195--202},
 booktitle = {4th fib International Conference on Concrete Sustainability (ICCS2024)},
 year = {2025},
}

@book{Saltelli.2008,
 author = {Saltelli, A. and Ratto, M. and Andres, T.},
 
 year = {2008},
 title = {Global sensitivity analysis: {T}he {P}rimer},
 address = {Chichester, England},
 publisher = {{Wiley-Interscience}}
}

@article{Taube.2025,
 author = {Taube, C. and Flohr, A. and Morgenthal, G.},
 year = {2025},
 title = {Experimental investigations on the influence of the short-term load history on the creep behaviour of normal concrete},
 pages = {140916},
 volume = {473},
 issn = {09500618},
 journal = {Construction and Building Materials},
 doi = {10.1016/j.conbuildmat.2025.140916}
}

@inproceedings{Taube.2025b,
 author = {Taube, C. and Flohr, A. and Morgenthal, G. and Osburg, A.},
 title = {Experimental Investigations on the Influence of Tailored Preloading on the Creep Behaviour of Concrete and its Variability},
 pages = {71--79},
 booktitle = {4th fib International Conference on Concrete Sustainability (ICCS2024)},
 year = {2025},

}

\end{document}